\documentclass[lettersize,journal]{IEEEtran}
\usepackage{amsmath,amsfonts}
\usepackage{algorithmic}
\usepackage{algorithm}
\usepackage{array}
\usepackage[caption=false,font=normalsize,labelfont=sf,textfont=sf]{subfig}
\usepackage{textcomp}
\usepackage{stfloats}
\usepackage{url}
\usepackage{verbatim}
\usepackage{graphicx}
\usepackage{cite}
\usepackage{color}
\usepackage{multirow}
\usepackage{multicol} 
\usepackage{mathptmx}

\DeclareMathAlphabet{\mathcal}{OMS}{cmsy}{m}{n}
\DeclareSymbolFont{largesymbols}{OMX}{cmex}{m}{n}

\begin{document}
\title{Heterogeneous Feature Augmentation for Ponzi Detection in Ethereum}

\author{Chengxiang Jin, Jie Jin, Jiajun Zhou, Jiajing Wu,~\IEEEmembership{Senior Member,~IEEE,} Qi Xuan,~\IEEEmembership{Senior Member,~IEEE}
\thanks{This work was partially supported by the National Key R\&D Program of China under Grant 2020YFB1006104, by the Key R\&D Programs of Zhejiang under Grants 2022C01018 and 2021C01117, by the National Natural Science Foundation of China under Grant 61973273, and by the Zhejiang Provincial Natural Science Foundation of China under Grant LR19F030001. \emph{(Corresponding author: Jiajun Zhou.)}}
\thanks{C. Jin, J. Jin, J. Zhou, Q. Xuan are with the Institute of Cyberspace Security, College of Information Engineering, Zhejiang University of Technology, Hangzhou 310023, China. E-mail:\{2112103081, 2112003197, jjzhou, xuanqi\}@zjut.edu.cn.}
\thanks{J. Wu is with the School of Computer Science and Engineering, Sun Yat-sen University, Guangzhou 510006, China. E-mail: wujiajing@mail.sysu.edu.cn.}
}

\markboth{Journal of \LaTeX\ Class Files,~Vol.~14, No.~8, August~2021}%
{Shell \MakeLowercase{\textit{et al.}}: A Sample Article Using IEEEtran.cls for IEEE Journals}


\maketitle

\begin{abstract}
  While blockchain technology triggers new industrial and technological revolutions, it also brings new challenges. 
  Recently, a large number of new scams with a "blockchain" sock-puppet continue to emerge, such as Ponzi schemes, money laundering, etc., seriously threatening financial security. 
  Existing fraud detection methods in blockchain mainly concentrate on manual feature and graph analytics, which first construct a homogeneous transaction graph using partial blockchain data and then use graph analytics to detect anomaly, resulting in a loss of pattern information. 
  In this paper, we mainly focus on Ponzi scheme detection and propose \textit{HFAug}, a generic Heterogeneous Feature Augmentation module that can capture the heterogeneous information associated with account behavior patterns and can be combined with existing Ponzi detection methods. 
  \textit{HFAug} learns the metapath-based behavior characteristics in an auxiliary heterogeneous interaction graph, and aggregates the heterogeneous features to corresponding account nodes in the homogeneous one where the Ponzi detection methods are performed. 
  Comprehensive experimental results demonstrate that our \textit{HFAug} can help existing Ponzi detection methods achieve significant performance improvement on Ethereum datasets, suggesting the effectiveness of heterogeneous information on detecting Ponzi schemes.
\end{abstract}


\section{Introduction}

\IEEEPARstart{B}{lockchain} is best known for its crucial applications in financial cryptocurrency platforms such as Ethereum. 
According to CoinMarketCap, as of January 2022, the total value of all digital currencies hits a new high of 2.27 trillion dollars.
However, the huge economic value of digital currency also makes it a target for cybercriminals, resulting in a large number of illegal activities such as Ponzi schemes, money laundering, phish scams, etc.
The popularity of digital currency allows criminals to find new ways to transfer funds, bringing Ponzi schemes, an offline fraud that originated 150 years ago, into the digital world.
Ponzi scheme~\cite{artzrouni2009mathematics} is a type of financial fraud disguised as ``high-yield'' investment programs, which use the money of new investors to pay interest and short-term returns to old investors for creating the illusion of profitability and then defraud more investments. 
One study~\cite{vasek2015there} estimates that Ponzi schemes operated through Bitcoin have collected more than 7 million dollars from September 2013 to September 2014.  
Therefore, understanding the behavior of Ponzi schemes and detecting them from cryptocurrency platforms would be crucial to maintaining the stability of the investment environment in the financial market.

There exists plenty of related work to model complex transaction networks~\cite{tao2021complex,lin2020modeling} for detecting Ponzi schemes.
Massimo et al.~\cite{bartoletti2018data} collected rich real data through multi-input heuristic address clustering and extracted the most discriminating features associated with Ponzi schemes. 
Chen et al.~\cite{chen2018detecting,chen2019exploiting} proposed a machine learning-based Ponzi scheme identification method that focuses on analyzing the characteristics of contract transactions and counting contract byte codes. 
Fan et al.~\cite{fan2020expose} improved a combination of feature engineering and machine learning by training a Ponzi detection model using the idea of ordered augmentation. 
Wang et al.~\cite{wang2021ponzi} considered contract account characteristics and contract code characteristics, and used LSTM to recognize Ponzi. 
What's more, Chen et al.~\cite{chen2021improving} generated word embedding based on smart contract source code, and used multi-channel TextCNN and Transformer to automatically learn code features. 
Yu et al.~\cite{yu2021ponzi} first constructed the initial features for Ethereum accounts via manual feature engineering, and then updated account features using GCN~\cite{kipf2016semi}, finally detected Ponzi schemes.
Zhang et al.~\cite{zhang2021detecting} extracted the bytecode feature and mixed it with transaction and opcode frequencies, and then used the LightGBM to identify Ponzi schemes.

These above-mentioned Ponzi detection methods are mainly combined with several graph-related algorithms.
DeepWalk~\cite{perozzi2014deepwalk} and Node2Vec~\cite{grover2016node2vec} utilize random walks to obtain sequences of nodes, and then use skip-gram models to learn the representation of nodes to predict their structural information in homogeneous networks. 
Among GNNs, apart from the GCN method mentioned above, GIN~\cite{xu2018powerful} is proposed to make GNNs applicable to different graph structures, which is as powerful as the WL test in terms of prejudiced power and expressiveness. 
GraphSAGE~\cite{hamilton2017inductive} samples neighboring nodes based on GCN, and trains different aggregation functions to obtain a more accurate representation of the new nodes.


\begin{figure}[htp]
    \centering
    \includegraphics[width=\linewidth]{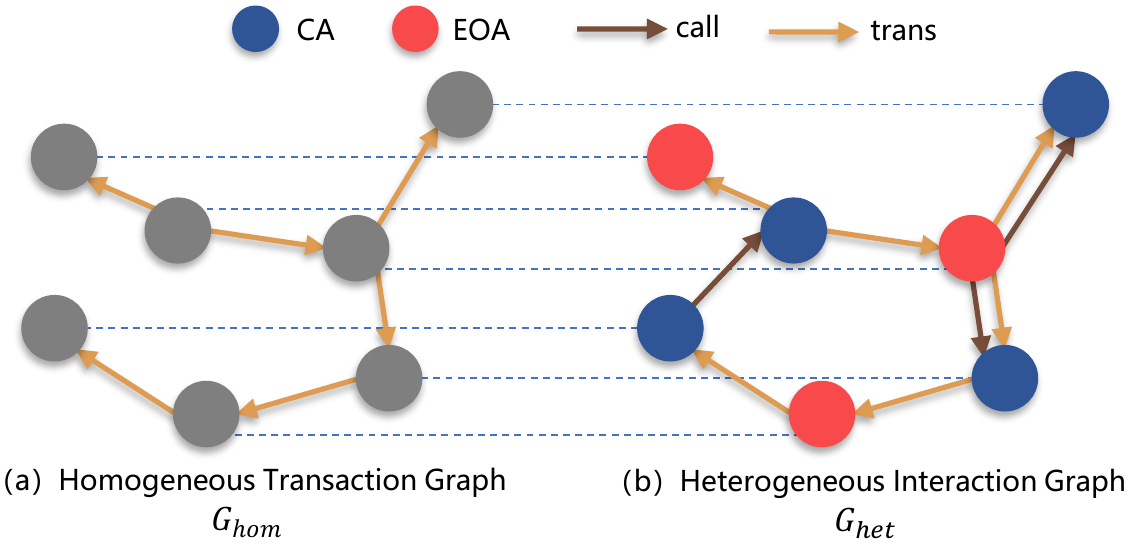}
    \caption{Homogeneous transaction graph and heterogeneous interaction graph.}
    \label{fig:hom_het}
\end{figure}
However, existing methods suffer from several shortcomings. 
Manual feature engineering usually designs statistical features related to transaction amount and time, but has difficulty defining more complex features that reflect transaction behavior.
Graph analytics is usually performed on a simple homogeneous transaction graph, failing in capturing the structural features associated with specific behavior patterns.
In this paper, we mainly focus on detecting Ponzi schemes on Ethereum, and consider improving the feature utilization of blockchain data and propose \textit{HFAug}, a generic \textbf{H}eterogeneous \textbf{F}eature \textbf{Aug}mentation module that can be adapted to various existing Ponzi detection methods.
\textit{HFAug} first extracts the metapath-based features on an auxiliary heterogeneous graph where the coordinated transaction and contract call information contained, and then aggregates these heterogeneous features associated with behavior patterns to corresponding account nodes in the homogeneous graph where the Ponzi detection methods are performed.
Our proposed module allows for improving the performance of existing Ponzi detection methods through feature augmentation without adjusting them.

The main contributions of this work are summarized as follows:
\begin{itemize}
\item [$\bullet$]We collect the labeled data of Ethereum Ponzi schemes for Ponzi detection research, and construct homogeneous transaction graph and heterogeneous interaction graph.
\item [$\bullet$]We propose a generic heterogeneous feature augmentation module, named \textit{HFAug}, which allows for aggregating heterogeneous features associated with behavior patterns to homogeneous transaction graphs, further improving the performance of existing Ponzi detection methods.
To the best of our knowledge, there are hardly any heterogeneous algorithms applied to blockchain data mining, and our work earlier explored the heterogeneous strategies for Ethereum Ponzi detection.
\item [$\bullet$]Extensive experiments on the Ethereum dataset show the effectiveness of \textit{HFAug} module on improving the performance of three categories of existing Ponzi detection methods. Moreover, the generic compatibility of \textit{HFAug} also suggests that heterogeneous behavior pattern information can benefit Ponzi scheme detection in Ethereum.
\end{itemize}

\section{Account Interaction Graph Modeling} \label{sec: graph-modeling}
\begin{figure}[htp]
    \centering
    \includegraphics[width=\linewidth]{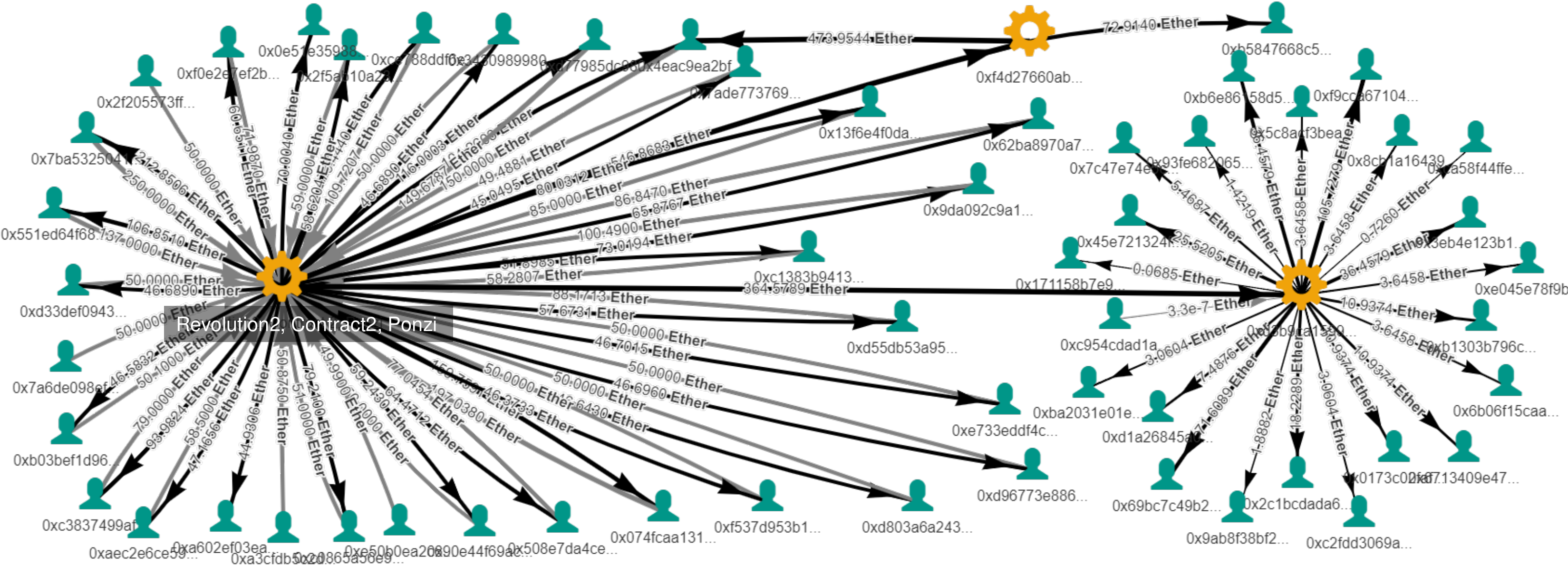}
    \caption{An interaction graph of real Ponzi scheme.} 
    \label{fig: ponzi}
\end{figure} 
\subsection{Ethereum Data}
An \emph{account} in Ethereum is an entity that owns Ether, and can be divided into two categories: Externally Owned Account (EOA) and Contract Account (CA).
EOA is controlled by a user with the private key and can initiate transactions on Ethereum, and CA is controlled by smart contract code and can only send transactions in response to receiving a transaction.
There are generally two categories of interactions between Ethereum accounts: \emph{transaction} and \emph{contract call}.
The \emph{transaction} refers to an action initiated by an EOA, and can be received by EOA or CA.
The \emph{contract call} refers to the process of triggering smart contract codes which can execute many different actions, such as transferring tokens or even creating a new contract.

\subsection{Graph Modeling on Ethereum}

\subsubsection{Homogeneous and Heterogeneous Graph}
The existing Ponzi detection methods usually model Ethereum data as a homogeneous graph, where all accounts will be treated as nodes of the same type, and interactions involving only transaction amounts will be treated as edges. 
Different from it, heterogeneous graph with different types of nodes and edges will retain more information of Ethereum data. 
More formally, we use $G_\textit{hom}=(V, E, Y)$ and $G_\textit{het}=(V_\textit{eoa}, V_\textit{ca}, E_\textit{trans}, E_\textit{call}, Y)$ to represent the two types of graph respectively, where $V$ represents the set of arbitrary accounts in the Ethereum data, $E$ represents the set of directed edges constructed from transaction information, $Y=\{(v_\textit{i}^\textit{p}, y_\textit{i})\}$ is the label information of known Ponzi accounts. 
Notably, all the known Ponzi schemes we have collected on Ethereum are based on contract accounts.

The nodes of $G_\textit{hom}$ and $G_\textit{het}$ are aligned, as illustrated in Fig.~\ref{fig:hom_het}.
Compared with $G_\textit{hom}$, $G_\textit{het}$ has additional account category information (i.e., EOA and CA), and another interactive edge information (i.e., contract call).

\subsubsection{Node Feature Construction}~\label{sec: node-feature}
We construct initial features for account nodes in both $G_\textit{hom}$ and $G_\textit{het}$ using 15 manual features proposed in existing methods.
\begin{itemize}
    \item [$\bullet$] The income and expenditure of the target account (including total, average, maximum and variance).
    \item [$\bullet$] The expenditure-income ratio of the target account.
    \item [$\bullet$] The balance of the target account. 
    \item [$\bullet$] The number of transactions sent and received by the target account.
    \item [$\bullet$] The investment Gini and return Gini of the target account.
    \item [$\bullet$] The life cycle of the target account.
\end{itemize}

\begin{figure*}[ht]
    \centering
    \includegraphics[width=\textwidth]{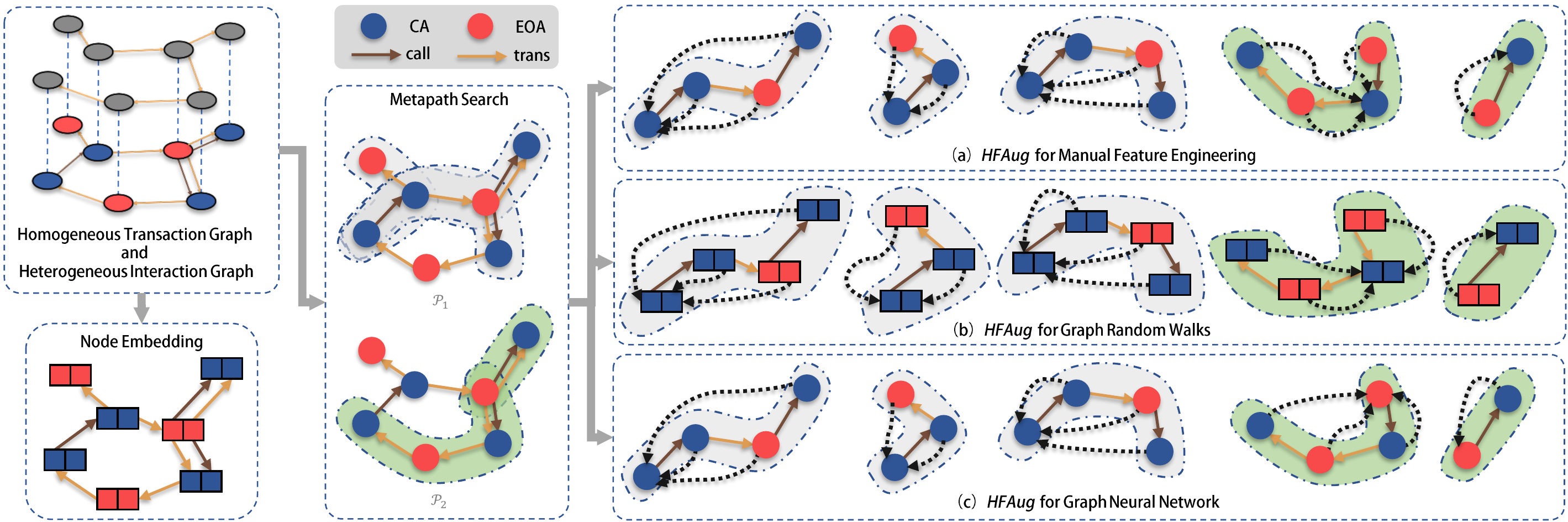}
    \caption{Illustration of Heterogeneous Feature Augmentation (\textit{HFAug}) module.} 
    \label{fig: framework}
\end{figure*} 

\subsubsection{Metapath}
Metapath~\cite{sun2012mining} is a path in a heterogeneous graph that contains a sequence of relations defined between different types of objects.
According to the interaction graph of Ponzi schemes, as schematically depicted in Fig.~\ref{fig: ponzi}, we predefine the critical behavior patterns as follows:
\begin{equation}\label{eq: ponzi-mode}
    EOA_1 \stackrel{call}{\longrightarrow}  CA_t \left(\stackrel{call}{\longrightarrow} CA_1 \right)\stackrel{trans}{\longrightarrow} EOA_2 \stackrel{call/trans}{\longrightarrow } CA_2 .
\end{equation}
External investors $EOA_1$ will transfer Ether to the Ponzi account $CA_\textit{t}$, which would perform subsequent actions.
$EOA_2$ could be an external investor or the Ponzi contract creator. The former indicates that \emph{trans} is a payback, while the latter indicates that \emph{trans} is a funds transfer.
Notably, very few Ponzi accounts will trigger internal calls ($\stackrel{call}{\longrightarrow} CA_1$) to perform subsequent actions.

We then extract two metapaths from above behavior patterns:
\begin{equation}
    \begin{array}{l}
        \mathcal{P}_{1}: CA_t \stackrel{call}{\longrightarrow} CA \stackrel{trans}{\longrightarrow} EOA \stackrel{call}{\longrightarrow} CA, \\
        \mathcal{P}_{2}: EOA \stackrel{call}{\longrightarrow} CA_t \stackrel{trans}{\longrightarrow} EOA \stackrel{trans}{\longrightarrow} CA.
    \end{array}
\end{equation}
Our \textit{HFAug} will caputre the behavior features from $G_\textit{het}$ based on these metapaths, as detailedly described below.

\section{Methodology} \label{sec:Methodology}
\subsection{\textit{HFAug} for Manual Feature and Graph Random Walks} \label{sec:4.1}
\subsubsection{Original Ponzi Detection}
For Ponzi detection methods based on manual feature engineering, we use the 15 manual features mentioned in Sec.~\ref{sec: node-feature} to characterize these CA, yielding the feature matrix $\mathbf{X} \in \mathbb{R}^{n\times 15}$, where $n$ represents the number of CA to be detected.
As for methods based on graph random walks, we generate structural embeddings as account node features rather than the predefined manual featres.
After that, the initial feature of arbitrary account node $v_\textit{i}$ is denoted as follows:
\begin{equation}
    \mathbf{x}_\textit{i}=\left\{\begin{array}{ll}
        [x_\textit{i}^1, x_\textit{i}^2, \cdots, x_\textit{i}^{15}] & \text {\textbf{for} manual feature} \vspace{5pt}\\
        {Walk}\left(G_\textit{hom}, v_i\right) & \text {\textbf{for} graph random walks}
        \end{array}\right.
\end{equation}
Finally, we achieve Ponzi detection by feeding account features into machine learning classifiers.

\subsubsection{Detection with \textit{HFAug}}

Here, \textit{HFAug} module is used to update the initial node features, as illustrated in Fig.~\ref{fig: framework}(a) and (b). 
Specifically, for a target CA node $v_\textit{ca}^\textit{t}$, we first search the target metapaths $\mathcal{P}_1$ or $\mathcal{P}_2$ where it is located in $G_\textit{het}$.
Notably, in $\mathcal{P}_1$ and $\mathcal{P}_2$, the $CA_\textit{t}$ is the target CA node.
After getting the metapath $\mathcal{P}$, we update the features of the target CA in $G_\textit{hom}$ by aggregating the features of other nodes in the metapath to it.
When the full metapath is not available, we only aggregate node features in the available subset of it.
The process of feature update can be represented as follows:
\begin{equation}
    \bar{\mathbf{x}}_\textit{ca}^\textit{t} =  \sum_{v \in \mathcal{P}^\prime \in \mathcal{P} } \textbf{x}_v,
\end{equation}
where $\mathbf{x}$ is the account feature, and $\mathcal{P}^\prime$ is the target metapath or its subset.

Finally, the updated features $\bar{\mathbf{X}}$ contain heterogeneous structural information associated with behavior patterns, and will be used for detecting Ponzi accounts.

\subsection{\textit{HFAug} for Graph Neural Network}
\subsubsection{Original Ponzi Detection}
GNN-based methods usually consider Ponzi detection as a node classification task.
In this paper, we consider three commonly used GNN models: GCN, GraphSAGE and GIN.
During Ponzi detection, the input is the homogeneous transaction graph $G_\textit{hom}$, and the output is a prediction of whether the target account is a Ponzi account.
The initial node features are also constructed according to Sec.~\ref{sec: node-feature}.

\subsubsection{Detection with \textit{HFAug}}
Here, \textit{HFAug} module is used to update the initial node features, as illustrated in Fig.~\ref{fig: framework}(c).
Specifically, these two metapaths are used to update the features of their respective head nodes.
In other words, for a CA/EOA node $v_\textit{ca}^\textit{t}$/$v_\textit{eoa}^\textit{t}$, we search the target metapath $\mathcal{P}_1$/$\mathcal{P}_2$ where it serves as the head node in $G_\textit{het}$, and update the features of head node in $G_\textit{hom}$ by aggregating the features of other nodes in the metapath to it.
The process of feature update can be represented as follows:
\begin{equation}
    \bar{\mathbf{x}}^\textit{t} =  \left\{\begin{array}{rll}
        \sum_{v \in \mathcal{P}^\prime \subseteq \mathcal{P}_1 } \textbf{x}_v, & \quad \mathcal{P}^\prime   \text { start from } v_\textit{ca}^\textit{t}, \\
        \sum_{v \in \mathcal{P}^\prime \subseteq \mathcal{P}_2 } \textbf{x}_v, & \quad \mathcal{P}^\prime   \text { start from } v_\textit{eoa}^\textit{t}.
        \end{array}\right.
\end{equation}
Notably, we can update one type of nodes using one metapath individually, or update all nodes using both metapaths simultaneously.

\section{Experiments} \label{sec:4}
\subsection{Data}

\begin{table}[htp]
    \centering
    \caption{Statistics of the homogeneous and heterogeneous graphs. 
            $|V|$ and $|E|$ are the total number of nodes and edges respectively. 
            $|V_\textit{ca}|$ and $|V_\textit{eoa}|$ are the number of CA and EOA respectively, 
            $|E_\textit{call}|$ and  $|E_\textit{trans}|$ are the number of call and trans edges respectively,
            and $|Y|$ is the number of labeled Ponzi accounts.}
    \resizebox{\linewidth}{!}{
    \renewcommand\arraystretch{1.1}
    \begin{tabular}{lccccccc} 
    \hline\hline
        Dataset                        & $|V|$  & $|E|$     & $|V_\textit{ca}|$ & $|V_\textit{eoa}|$ & $|E_\textit{call}|$ & $|E_\textit{trans}|$ & $|Y|$       \\ 
        \hline
        Homogeneous $G_\textit{hom}$   & 57,130 & 86,602    & \multicolumn{4}{c}{$\cdots \quad \text{No label information} \quad \cdots$}                      & 191                   \\
        Heterogeneous $G_\textit{het}$ & 57,130 & 156,255   & 4,616             & 52,514             & 69,653               & 86,602              & 191              \\
    \hline\hline
    \end{tabular}}
    \label{tb: data}
\end{table}

\begin{table}[htp]
    \centering
    \caption{Ponzi detection results of raw methods (manual feature engineering and random walk-based graph embedding) and their enhanced versions (with \textit{HFAug}).
    \textit{gain} represents the relative improvement rate.    
    }
    \label{tb: res-1}
    \resizebox{\linewidth}{!}{
    \renewcommand\arraystretch{1.1}
    \begin{tabular}{clccc|ccc} 
    \hline\hline
    \multicolumn{2}{c}{\multirow{2}{*}{Methods}}                                                             & \multicolumn{3}{c|}{$\mathcal{P}_1$}                                & \multicolumn{3}{c}{$\mathcal{P}_2$}                            \\ 
    \cline{3-8}
    \multicolumn{2}{c}{}                                                                                     & \multicolumn{1}{c}{LR} & SVM              & \multicolumn{1}{c|}{RF} & \multicolumn{1}{c}{LR} & SVM              & RF                \\ 
    \hline
    \multirow{3}{*}{\begin{tabular}[c]{@{}c@{}}Manual\\Feature\end{tabular}} & \textit{raw}                  & 65.73                  & 72.79            & 77.23                  & 65.73                   & 72.79            & 77.23             \\
                                                                            & \textit{raw} + \textit{HFAug} & 71.72                  & 76.18            & 74.61                  & 75.12                   & 76.96            & 75.65             \\
                                                                            & \textit{gain}                 & \textbf{\scriptsize{+9.11\%}}       & \textbf{\scriptsize{+4.66\%}} & \scriptsize{-3.39\%}               & \textbf{\scriptsize{+14.30\%}}       & \textbf{\scriptsize{+5.73\%}} & \scriptsize{-2.05\%}           \\ 
    \hline
    \multirow{3}{*}{DeepWalk}                                                & \textit{raw}                  & 80.63                  & 82.98            & 82.74                  & 80.63                   & 82.98            & 82.74             \\
                                                                            & \textit{raw} + \textit{HFAug} & 81.43                  & 84.58            & 81.43                  & 80.64                   & 81.95            & 83.26             \\
                                                                            & \textit{gain}                 & \textbf{\scriptsize{+0.99\%}}       & \textbf{\scriptsize{+0.19\%}} & \scriptsize{-0.02\%}       & \textbf{\scriptsize{+0.00\%}}       & \scriptsize{-1.24\%}          & \textbf{\scriptsize{+0.63\%}}  \\ 
    \hline
    \multirow{3}{*}{Node2Vec}                                                & \textit{raw}                  & 82.22                  & 84.56            & 86.14                  & 82.22                   & 84.56            & 86.14             \\
                                                                            & \textit{raw} + \textit{HFAug} & 83.78                  & 86.93            & 86.67                  & 81.69                   & 84.83            & 86.14             \\
                                                                            & \textit{gain}                 & \textbf{\scriptsize{+1.90\%}}      & \textbf{\scriptsize{+2.80\%}} & \textbf{\scriptsize{+0.62\%}}       & \scriptsize{-0.64\%}                 & \textbf{\scriptsize{+0.32\%}} & \scriptsize{+0.00\%}  \\
    \hline\hline
    \end{tabular}}
\end{table}

\begin{table}[htp]
    \centering
    \caption{Ponzi detection results of raw methods (GNN-based methods) and their enhanced versions (with \textit{HFAug}). 
    + \textit{HFAug}($\mathcal{P}$) represents the results of detection methods enhanced by \textit{HFAug} with metapath $\mathcal{P}$.
    }
    \label{tb: res-2}
    \resizebox{\linewidth}{!}{%
    \renewcommand{\arraystretch}{1.1}
    \begin{tabular}{ccccc} 
    \hline\hline
    \begin{tabular}[c]{@{}c@{}}Methods\end{tabular} & \textit{raw}  & 
    \begin{tabular}[c]{@{}c@{}}+\textit{HFAug}\\$({\mathcal{P}_1})$\end{tabular} &
    \begin{tabular}[c]{@{}c@{}}+\textit{HFAug}\\$({\mathcal{P}_2})$\end{tabular} & 
    \begin{tabular}[c]{@{}c@{}}+\textit{HFAug}\\$({\mathcal{P}_1,\mathcal{P}_2})$\end{tabular}              \\ 
    \hline
    GCN              & 82.48 & \begin{tabular}[c]{@{}c@{}}82.66\\\scriptsize{+0.22\%}\end{tabular} & \begin{tabular}[c]{@{}c@{}}83.03\\\textbf{\scriptsize{+0.67\%}}\end{tabular} & \begin{tabular}[c]{@{}c@{}}\textbf{84.05}\\\textbf{\scriptsize{+1.90\%}}\end{tabular}  \\ 
    \hline
    GraphSAGE        & 78.54 & \begin{tabular}[c]{@{}c@{}}74.86\\\scriptsize{-4.68\%}\end{tabular} & \begin{tabular}[c]{@{}c@{}}78.61\\\textbf{\scriptsize{+0.10\%}}\end{tabular} & \begin{tabular}[c]{@{}c@{}}\textbf{78.70}\\\textbf{\scriptsize{+0.20\%}}\end{tabular}  \\ 
    \hline
    GIN              & 77.59 & \begin{tabular}[c]{@{}c@{}}77.50\\\scriptsize{-0.11\%}\end{tabular} & \begin{tabular}[c]{@{}c@{}}77.93\\\textbf{\scriptsize{+0.44\%}}\end{tabular} & \begin{tabular}[c]{@{}c@{}}\textbf{78.05}\\\textbf{\scriptsize{+0.60\%}}\end{tabular}  \\
    \hline\hline
    \end{tabular}}
\end{table}

We collected 191 labeled Ponzi data from \textit{Xblock}\footnote[1]{\url{http://xblock.pro/ethereum/}}, \textit{Etheriscan}\footnote[2]{\url{https://cn.etherscan.com/accounts/label/ponzi}} and other Blockchain platforms.
For all detection methods, we take all the labeled Ponzi accounts as positive samples, as well as the same number of randomly sampled CA as negative samples.
We construct the homogeneous transaction graph using the transaction data of these CA, yielding a graph with 56,748 nodes and 86,602 edges.
For the heterogeneous interaction graph, we divide all the nodes into two categories: 4,616 CA and 52,514 EOA, and add additional 69,653 call edges.
The statistics of data are shown in Table~\ref{tb: data}.

\subsection{Ponzi Detection Methods and Experimental Setup}
To illustrate the effectiveness of our \textit{HFAug} module, we combine it with three categories of Ponzi detection methods: manual feature engineering, random walk-based graph embedding and GNN-based methods.

For manual feature engineering which is the most common and simplest method for Ponzi detection, we use 15 manual features listed in Sec~\ref{sec: node-feature}, yielding account feature vectors with dimension equals to 15.
For random walk-based graph embedding, we consider DeepWalk and Node2Vec.
For the above two categories of methods, we achieve Ponzi detection by feeding the generated account features into three machine learning classifiers: Logistic Regression (LR), Support Vector Machine (SVM) and Random Forest (RF).
For GNN-based methods, we compare with three commonly used GNNs: GCN, GraphSAGE and GIN.

For walk-based methods, we set the dimension of embedding, window size, walk length and the number of walks per node to 128, 10, 50 and 5 respectively. 
For Node2Vec, we perform a grid search of return parameter $p$ and in-out parameter $q$ in \{0.5, 1, 2\}.
For GNN-based methods, we set the hidden dimension of GCN, GrahphSAGE and GIN to 128, 512 and 128 respectively, and the learning rate to 0.005, 0.001 and 0.01 respectively.
For all methods, we recept 5-fold cross validation 10 times and report the average micro-F1 score.

\subsection{Evaluation}
We evaluate the benefit of our \textit{HFAug} on enhancing Ponzi detection, answering the following research questions:
\begin{itemize}
    \item [$\bullet$] \textbf{RQ1}: Can \textit{HFAug} improve the performance of Ponzi detection when being combined with existing detection methods?
    \item [$\bullet$] \textbf{RQ2}: Whether the enhancement effect of \textit{HFAug} is determined by the extracted heterogeneous information?
\end{itemize}
We combine the proposed heterogeneous feature augmentation module with all Ponzi detection models to show a crosswise comparison.

\subsubsection{Enhancement for Ponzi Detection}
Table~\ref{tb: res-1} and \ref{tb: res-2} report the results of performance comparison between the raw methods and their enhanced version (with \textit{HFAug}), from which we observe that there is a significant boost in detection performance across all methods.
Overall, these detection methods combined with \textit{HFAug} module obtain higher average detection performance in most cases, and the \textit{HFAug} achieves a 70.37\% success rate\footnote{The success rate refers to the percentage of enhanced methods with F1 score higher than that of the corresponding raw methods in Table~\ref{tb: res-1} and \ref{tb: res-2}.} on the enhancement of Ponzi detection.

Specifically, for manual feature engineering, we observe 4.66\% $\sim$ 14.30\% relative improvement on LR and SVM classifiers, as well as a negative gain for RF classifier.
It is obvious that manual features have poor expressiveness compared to other methods and heavily relies on the performance of classifiers.
We speculate that our \textit{HFAug} has a better enhancement for manual feature engineering with weak classifiers.
For the walk-based methods and GNN-based methods, the learnt features are better at capturing the behavior patterns of accounts than manual features, manifesting as higher raw performance.
For both types of methods, the module achieves a relatively limited boost.

These phenomena provide a positive answer to \textbf{RQ1}, indicating that the \textit{HFAug} module can benefit the existing Ponzi detection methods via feature augmentation and improve their performance without adjusting them.

\subsubsection{Impact of Metapaths}\label{sec: impact-metapath}
We further investigate the influence of metapaths in \textit{HFAug} on the enhancement effect.
As we can see from Table~\ref{tb: res-1} and \ref{tb: res-2}, \textit{HFAug} with $\mathcal{P}_2$ outperforms that with $\mathcal{P}_1$ in most cases, which suggesting that the performance of \textit{HFAug} relies on the choice of metapaths.

Both $\mathcal{P}_1$ and $\mathcal{P}_2$ are extracted from the basic behavior patterns of Ponzi scheme defined in Eq.~\ref{eq: ponzi-mode}, and we have reasonable explanations for their performance difference: 
1) Fewer metapath in heterogeneous interaction graph start with CA than EOA;
2) Ponzi contracts usually have more frequent interactions with EOA;
3) metapath $\mathcal{P}_1$ contains the behavior of internal calls (i.e., $CA \stackrel{call}{\longrightarrow} CA $), which is relatively rare.

For manual feature engineering and GNN-based methods, we use the manual feature rather than embedding as initial node feature, which does not contain additional structural information. 
As a result, metapath $\mathcal{P}_2$ which reflects more frequent behavior patterns reasonably achieves superior performance compared with $\mathcal{P}_1$.
For walk-based methods, the result with metapath $\mathcal{P}_2$ is not better than $\mathcal{P}_1$, and we make the following reasonable explanations.
Combine the following two prior knowledge: 1) metapath $\mathcal{P}_1$ starts from the target node while metapath $\mathcal{P}_2$ not, and 2) the embedding of the target node is generated from the walks starting from the target node, we speculate that metapath $\mathcal{P}_2$ updates the feature of target node by aggregating the information along the metapath, including the head node $EOA$ that has a high probability of not appearing in the walks, which may lead to a conflict between the heterogeneous information defined by the metapath and the structural information learned by the random walks, further bringing poor performance.

Furthermore, we observe that a combination of multiple metapaths can perform better than a single metapath, as shown in Table~\ref{tb: res-2}, suggesting that multiple heterogeneous information can benefit Ponzi detection more.
These phenomena provide a positive answer to \textbf{RQ2}, indicating that the design of the metapaths is critical and determines whether the \textit{HFAug} can effectively capture the heterogeneous information associated with account behavior patterns.



\section{Conclusion} \label{sec:5}
Existing Ponzi detection methods usually ignore the structural behavior patterns of Ponzi accounts, resulting in a loss of information.
In this paper, we propose a generic Heterogeneous Feature Augmentation module which can capture the heterogeneous information associated with account behavior patterns and can be combined with existing Ponzi detection methods.
Comprehensive experiments show that our
\textit{HFAug} can help existing Ponzi detection methods achieve significant improvement on Ethereum datasets.
Moreover, we also conclude that the enhancement effect of \textit{HFAug} is determined by the extracted heterogeneous information, which encourages us to design more highly-expressive metapaths in future work.



\bibliographystyle{IEEEtran}
\bibliography{ref}
\end{document}